\documentclass[oldversion]{aa}

\usepackage{graphicx}
\usepackage{natbib}
\usepackage{textcomp}
\usepackage{amsmath}
\usepackage{amssymb}
\usepackage{wasysym}
\usepackage{fancybox}
\usepackage{multirow}

\newcommand{\un}[0]{\mbox{e$^{-}$/32s}}

\newcommand{\fe}[0]{ps}

\begin{document}

\title{How stellar activity affects the size estimates of extrasolar planets}

\author{S. Czesla
   \and K. F. Huber
   \and U. Wolter
   \and S. Schr\"oter
   \and J. H. M. M. Schmitt}

\institute{Hamburger Sternwarte, Universit\"at Hamburg, Gojenbergsweg 112, 21029 Hamburg, Germany}

\date{Received ... / Accepted ... }

           \abstract {Lightcurves have long been used to study stellar activity and have more recently become a major
           tool in the field of exoplanet research.
           We discuss the various ways in which stellar activity can influence transit lightcurves, and study
           the effects using
           the outstanding photometric data of the CoRoT-2 exoplanet system.
           We report a relation between the `global' lightcurve and the transit profiles which turn out to be
           shallower during high spot coverage on the stellar surface.
           Furthermore, our analysis reveals a color dependence of the transit lightcurve
           compatible with a wavelength-dependent limb darkening law as observed on the Sun.
           Taking into account activity-related effects, we re-determine the orbit inclination and planetary radius and find the
           planet to be $\approx 3$\% larger than reported previously.
           Our findings also show that exoplanet research cannot generally ignore the effects of stellar activity.
           }


\maketitle

%%%%%%%%%%%%%%%%%%%%%%%%%%%%%%%%%%%%%%%%%%%%%%%%%%%%%%%%
\section{Introduction}
\label{Sec:Intro}

The brightness distribution on the surface of active stars is both spatially inhomogeneous and temporally variable.
The state and evolution of the stellar surface structures can be traced by the rotational and secular modulation of the observed photometric lightcurve.
In the field of planet research, lightcurves including planetary transits are of particular interest, since
they hold a wealth of information on both the planet and its host star.

The outstanding quality of the space-based photometry provided by the CoRoT mission \citep[e.g.][]{Auvergne2009} provides stellar
lightcurves with an unprecedented precision, temporal cadence and coverage. While primarily designed 
as a planet finder, the CoRoT data are also extremely interesting in the context of
stellar activity.  Recently, \citet{Lanza2009} demonstrated the information content to be extracted from
such lightcurves in the specific case of CoRoT-2a. This star is solar-like in mass and radius, but
rotates faster at a speed of \mbox{$v\sin(i) \, = \, 11.85 \, \pm \, 0.50$ km/s} \citep{Bouchy2008}. Its rotation period
of \mbox{$\approx 4.52$~days} was deduced from slowly evolving active regions, which dominate the photometric variations.
Thus, CoRoT-2a is a very active star by all standards. Even more remarkably, CoRoT-2a is orbited by a giant
planet \citep{Alonso2008}, which basically acts as a shutter scanning the surface of CoRoT-2a along a well defined 
latitudinal band.

The transiting planetary companion provides a key to understanding the surface structure of its
host star. While previous
analyses have either ignored the transits \citep{Lanza2009} or the `global' lightcurve \citep{Wolter2009},
we show that there is a relation between the transit shape and the global lightcurve which cannot generally be neglected in
extrasolar planet research.

%%%%%%%%%%%%%%%%%%%%%%%%%%%%%%%%%%%%%%%%%%%%%%%%%%%%%%%%
\section{Observations and data reduction}
\label{Sec:Object}

\citet{Alonso2008} discovered the planet CoRoT-2b using the photometric CoRoT data (see Table~\ref{tab:Exo2prop}). Its host star
has a spectral type of G7V with an optical (stellar) companion too close to be resolved by CoRoT.
According to \citet{Alonso2008}, this secondary contributes a constant $(5.6 \, \pm 0.3)$\% of the total CoRoT-measured flux.
CoRoT-2b's orbital period of \mbox{$\approx 1.74$~days} is about one third of CoRoT-2a's rotation period, and the 
almost continuous CoRoT data span $142$~days, sampling about $30$~stellar rotations and more than  $80$~transits.
The lightcurve shows clear evidence of strong activity: There is substantial modulation
of the shape on time scales of several days, and the transit profiles are considerably deformed as a consequence of surface inhomogeneities
\citep{Wolter2009}.

\begin{table}[t!]
  \begin{minipage}[h]{0.5\textwidth}
    \renewcommand{\footnoterule}{}
    \caption{Stellar/planetary parameters of CoRoT-2a/b. \label{tab:Exo2prop}}
    \begin{center}
      \begin{tabular}{l c c}
      \hline \hline
      Star \hspace*{0.0cm} \footnote{$P_s$ - stellar rotation period.}  & Value $\pm$ Error & Ref.\footnote{taken from \citet{Lanza2009} [L09], \citet{Alonso2008} [A08], or \citet{Bouchy2008} [B08]} \\
      \hline
      $P_s$          & $(4.522 \, \pm \, 0.024)$ d         & L09 \\
      Spectral type  & G7V                                 & B08 \\
      \hline \hline
      Planet \hspace*{0.0cm} \footnote{$P_p$ - orbital period, $T_c$ - central time of first transit, $i$ - orbital inclination, $R_p, R_s$ - planetary and stellar radii, $a$ - semi major axis of planetary orbit, $u_a, u_b$ - linear and quadratic limb darkening coefficients.}  & Value $\pm$ Error & Ref.\\
      \hline
      $P_p$       & $(1.7429964 \, \pm \, 0.0000017)$ d   & A08 \\
      $T_c$ [BJD] & $(2454237.53362 \, \pm \, 0.00014)$ d & A08 \\
      $i$            & $(87.84 \, \pm \, 0.10)$\textdegree & A08 \\
      $R_p/R_s$   & $(0.1667 \, \pm \, 0.0006)$           & A08 \\
      $a/R_s$     & $(6.70 \, \pm \, 0.03)$               & A08 \\
      $u_a, u_b$      & $(0.41\pm0.03), (0.06\pm 0.03)$           & A08 \\
      \hline
      \end{tabular}
    \end{center}
  \end{minipage}
\end{table}

  \label{sec:DataReduction}
  Our data reduction starts with the results provided by the CoRoT~N2 pipeline (\texttt{N2\_VER}~$1.2$).
  CoRoT provides three band photometry (nominally red, green, and blue), which we extend by a virtual fourth
  band resulting from the combination (addition) of the other bands. This `white' band is, henceforth, treated as
  an independent channel, and our analysis will mainly refer to this band.
  It provides the highest count rates and, more importantly, is less susceptible to
  instrumental effects such as long term trends and `jumps' present in the individual color channels.
  
  In all bands we reject those data points
  flagged as `bad' by the standard CoRoT pipeline (mostly related to the South Atlantic anomaly).
  The last step leaves obvious outliers in the light curves. To remove them, we
  estimate the standard deviation of the data point distribution in short ($\approx 3000$~s) slices and reject
  the points more than $3\sigma$ off a (local) linear model.
  Inevitably, we also remove a fraction of physical data (statistical outliers) in this step, but we estimate
  that loss to be less than a percent of the total number of data points, which we consider acceptable.
  
  In all bands the white we find photometric discontinuities (jumps) which are caused by particle impact on
  the CoRoT detector. In the case of CoRoT-2a the jumps are of minor amplitude compared
  to the overall count rate level, and we correct them by adjusting the part of the lightcurve following the jump to the
  preceding level.
  
  Finally, we correct the CoRoT photometry for systematic, instrumental trends
  visible in all bands but white. In order to approximate the instrumental trend,
  we fit the (entire) lightcurve with a second order polynomial, $q$, and apply the equation
  \begin{equation}
    c_{corr,i} = c_{o, i} \cdot \frac{\overline{c}}{q_i} \; .
    \label{eq:LongTermCorr}
  \end{equation}
  Here, $c_{o, i}$ is the i-th observed data point and $q_i$ the associated value of the best fit second order polynomial,
  $\overline{c}$ represents the mean of all observed count rates in the band,
  and $c_{corr,i}$ the corrected photometry.
  
  The resulting lightcurve still shows a periodic signal clearly related to the orbital motion of the CoRoT satellite. Again,
  this is a minor effect in the white band, and we neglect this in the context of the following analysis.
  
  In a last step we subtract $5.6$\% of the median lightcurve level to account for the companion contribution. Note that we
  use the same rule for all bands which can only serve as an approximation as \citet{Alonso2008} point out that the companion has a later
  type (probably K or M) and, therefore, a different spectrum than CoRoT-2a.

\section{Analysis}
\label{Sec:Analysis}

\subsection{Transit profiles and stellar activity}
  A planet crossing the stellar disk imprints a characteristic transit feature on the lightcurve of the star \citep[e.g.][]{Pont2007, Wolter2009}.
  The exact profile is determined by planetary parameters as well as the structure of the stellar surface.
  A model which describes the transit profile must account for both.
  One of the key parameters of the surface model is the limb darkening law. The presence
  of limb darkening seriously complicates transit modeling,
  because it can considerably affect the transit profile, while it is hard to recover its characteristics from lightcurve
  analyses \citep[e.g.,][]{Winn2009}.
  
  Stellar activity adds yet another dimension of complexity to the problem, because
  a (potentially evolving) surface brightness distribution also affects the transit profiles. The local brightness on
  the surface can either be decreased by dark spots or increased by bright faculae compared to the undisturbed photosphere.
  Spots (or faculae) located within the eclipsed section of the stellar surface lead to a decrease (increase) in the transit depth,
  and the actual profile depends on the distribution of those structures across the planetary path.
  Spots and faculae located on the non-eclipsed section of the surface do not directly affect
  the transit profile but change the overall level of the lightcurve.
  As transit lightcurves are, however, usually normalized with respect to the count rate level immediately before and after
  the transit,
  the non-eclipsed spot contribution enters (or can enter) the resulting curve as a time-dependent modulation of
  the normalized transit depth.

\subsection{Transit lightcurve normalization}
  As mentioned above, the normalization may affect the shape of the transit profiles.
  We now discuss two normalization approaches and compare their effect on the transit profiles.
  Let $f_i$ be the measured flux in time bin $i$, $n_i$ an estimate of the count rate level without the transit
  (henceforth referred to as the `local continuum'), and $p$ a
  measure of the unspotted photospheric level in the lightcurve, i.e., the count rate obtained in the respective band,
  when the star shows a purely photospheric surface. Usually, the quantity
  \begin{equation}
    y_i=f_i/n_i
    \label{eq:yiNorm}
  \end{equation}
  is referred to as the `normalized flux'.
  
  Normalization according to Eq.~\ref{eq:yiNorm} can, however, result in variations of the transit lightcurve depth in
  response to non-uniform surface flux distributions as encountered on active stars.
  Assume a planet transits its host star twice. During the first transit the stellar surface remains free
  of spots, but during the second transit there is a large active region anywhere on the star not covered by the planetary
  disk (but visible). Consequently, the local continuum estimate, $n_i$, for the second transit is lower, and the
  normalized transit appears deeper, although it is exactly the same transit in absolute (non-normalized) numbers.
  
  To overcome this shortcoming, we define the alternative normalization
  \begin{equation}
    z_i=\frac{f_i-n_i}{p}+1\, .
    \label{eq:ziNorm}
  \end{equation}
  In both cases the transit lightcurve is normalized with respect to the local continuum either by division or by
  subtraction. The conceptual difference lies in the treatment of the local continuum level and how it enters
  the normalized transit lightcurve. Using Eq.~\ref{eq:ziNorm} the observed transit is shifted, normalized by a constant, and
  shifted again. While the scaling in this case remains the same for all transits, the scaling applied in 
  Eq.~\ref{eq:yiNorm} is a function of the local continuum.
  
  Following the above example, we assume the same transit is normalized using both Eq.~\ref{eq:yiNorm}
  and Eq.~\ref{eq:ziNorm}. To evaluate the differences between the approaches we consider the expression
  \begin{equation}
    \frac{z_i}{y_i} = \frac{(f_i-n_i)/p+1}{f_i/n_i} \ge 1 \ .
    \label{eq:ziOveryi}
  \end{equation}
  For $n_i=p$ Eq.~\ref{eq:ziOveryi} holds as a strict equality, i.e. both normalizations yield identical
  results. The inequality equates to true if $p>n_i$ and $n_i>f_i$. The first condition reflects the fact that the
  local continuum estimate should not exceed the photospheric lightcurve level, and the second one says
  that the lightcurve level is below the local continuum. The second condition is naturally fulfilled during
  a transit, and the first one is also met as long as faculae do not dominate over the dark spots during the transit. At least
  in the case of CoRoT-2a \citet{Lanza2009} find no evidence for a significant flux contribution due to faculae, so that we conclude that the normalized
  transit obtained using Eq.~\ref{eq:ziNorm} is
  always shallower than the one resulting from Eq.~\ref{eq:yiNorm} unless $n_i=p$, in which case
  the outcomes are equal.
  
  \subsubsection{Quantifying the normalization induced difference in transit depth}
   \label{sec:QuanTheEffect}

   Let us now study a single transit and consider data points covered by index set $j$, for which the term \mbox{$n_j-f_j$} reaches a 
   maximal value of $T_0$ at some index value $j=T$. At this position the normalization obtained from Eq.~\ref{eq:ziNorm} is 
   given by $z_T=(f_T-n_T)/p+1=-T_0/p+1$, whereas
   Eq.~\ref{eq:yiNorm} yields $y_T=f_T/n_T=(n_T-T_0)/n_T$. These values are now used to compare the transit depths provided by the
   two normalizations.
   Note that we assume here that the normalized depth is
   maximal at index $T$; this is always true for Eq.~\ref{eq:ziNorm} but not necessarily for Eq.~\ref{eq:yiNorm} which we regard
   a minor issue.
   Again we find $z_T=y_T$ if $n_T=p$. If, however,
   the local continuum estimate is given by $n_T \approx \alpha p$ ($\alpha \le 1$), the results differ
   by
   \begin{equation}
     z_T-y_T  = T_0 p^{-1} \left(\alpha^{-1}-1\right) \; .
     \label{eq:normDiff}
   \end{equation}

   Using the extreme values observed for CoRoT-2a ($\alpha \approx 0.96$, and $T_0 \approx 0.03\times p$),
   the right hand side of Eq.~\ref{eq:normDiff} yields
   $\approx 1.3\times 10^{-3}$ for the difference in transit depth, caused exclusively by applying
   two different normalization prescriptions.
  
  \subsubsection{Which normalization should be used?}
   For planetary research it is important to `clean' the transit lightcurves of stellar activity in order to find the
   `undisturbed' profile associated with the planet \textit{only}.
   As transit lightcurves normalized using Eq.~\ref{eq:ziNorm} are all scaled using the same factor, they preserve their
   shape and depth (at least relative to each other) and can, therefore, be combined consistently,
   which is not necessarily the case when Eq.~\ref{eq:yiNorm}
   is used.
   Note that this does not mean that the obtained transit depth is necessarily the `true' depth, because
   Eq.~\ref{eq:ziNorm} includes the photospheric brightness level, $p$, as a time independent scaling factor.
   At least in the context of the lightcurve analysis, $p$ cannot be determined with certainty as the star may
   not show an undisturbed surface during the observation; it may actually never show it.
   
   A problem evident in CoRoT lightcurve analyses is the
   existence of long term instrumental gradients in the data (cf. Sect.~\ref{sec:DataReduction}). Modeling these trends by
   a `sliding' response, $R_d$, of the detector so that the relation between `true' photometry, $c_i$, and observation, $c_{o,i}$, is given by
   $c_{i,o}=c_i\cdot R_{d,i}$, Eq.~\ref{eq:LongTermCorr} yields
   \begin{equation}
     c_{corr, i}=c_i\cdot \left( R_{d,i}\frac{\overline{c}}{q_i} \right) \, .
   \end{equation}
   Obviously, the true photometry is recovered when the embraced term equates to one. Yet the scaling of $\overline{c}$ in
   Eq.~\ref{eq:LongTermCorr} is arbitrary, so that this is not necessarily the case. As long as $q_i$, however, appropriately
   represents the shape of $R_{d,i}$ the term provides a global scaling which cancels out in both Eq.~\ref{eq:yiNorm} and
   Eq.~\ref{eq:ziNorm}.
   
   For our transit analysis we argue in favor of the normalization along Eq.~\ref{eq:ziNorm}.
   We estimate the photospheric level from the highest count rate during the most prominent global maximum
   (at JD$\approx 2454373.3$) in each individual band. These estimates are based on the reduced lightcurves, in particular, the
   instrumental trend and the stellar companion have been accounted for.
   Throughout our analysis we use the values
   $p_{white}=703\,000$, $p_{red}=489\,000$, $p_{green}=88\,500$, and $p_{blue}=124\,500$ (in units of~\un).
   Since even at that time spots are likely to be present on the stellar disk, these estimates might in fact represent lower limits to the
   true value of $p$.
   
\subsection{Transit profiles in CoRoT-2a}
  The global lightcurve of CoRoT-2a shows pronounced maxima and minima and a temporally variable amplitude of
  the global modulation \citep{Alonso2008}.
  It is natural to expect the spot coverage on the eclipsed section of the stellar surface
  to be smallest where the global lightcurve is found at a
  high level, and transit events occurring during those phases should, thus, be least contaminated with
  the effects of stellar activity.
  The contrary should be the case for transits during low lightcurve levels.
  
  To quantify the impact of activity on the transit profile, we define the transit equivalent width (TEW) of transit $n$
  \begin{equation}
    TEW_n = \int_{t_I}^{t_{IV}} \Big( 1-z_n(t) \Big)\, \mbox{d}t \approx \sum_i (1-z_{n,i}) \, \delta t_i \, ,
    \label{eq:TEW}
  \end{equation}
  where $t_I$ and $t_{IV}$ must be chosen so that they enclose the entire transit.
  Note that extending the integration boundaries beyond the actual extent of the transit does not change the
  expectation value of Eq.~\ref{eq:TEW}, but only introduces an extra amount of error. The nominal unit of
  the TEW is time.
  
  \subsubsection{A relation between transit equivalent width and global lightcurve modulation}
   As outlined above we expect a larger impact of activity where the overall lightcurve level is low. When this is true, it should be
   reflected by a relation between the transit equivalent width and the transit continuum level (the overall lightcurve level at transit time).
   
   \begin{figure}[h]
     \includegraphics[angle=-90, width=0.49\textwidth]{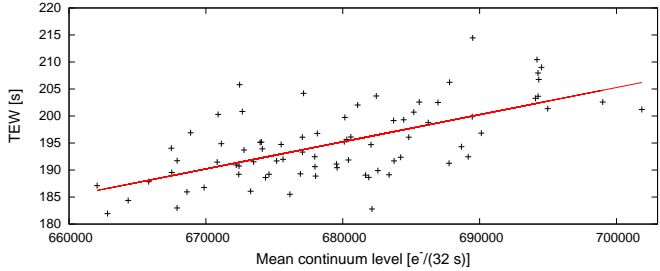}
     \caption{Transit equivalent width (TEW) vs. transit continuum level as well as the best-fit linear model.
     \label{fig:TEWvsCL}}
   \end{figure}
   
In Fig.~\ref{fig:TEWvsCL} we show the distribution of TEWs as a function of the local continuum level for all $79$ transits observed with a $32$~s sampling.
There is a clear tendency for larger TEWs to be associated with higher continuum levels, thus, providing
obvious evidence for activity-shaped transit lightcurves.
In the same figure, we also show the best fit for a linear model relation, which has a gradient of d(TEW)/d(CL)$=(5 \pm 1.5) \times 10^{-4}$ s/(e$^{-}32$ s).

To corroborate the reality of the above stated correlation, we calculated the correlation coefficient, $R$.
Its value of $R = 0.642$ confirms the visual impression of
a large scatter in the distribution of data points (cf., Fig.~\ref{fig:TEWvsCL}). We estimate the statistical error for
a single data point to be $\approx 0.1$\%, so that the scatter cannot be explained by measurement errors.
To check whether the continuum level and the TEW are
independent variables, we employ a t-test and find the null hypothesis (independent quantities) to be rejected with an error probability
of $1.8 \times 10^{-10}$, so that
the correlation between the TEWs and the continuum level must be regarded as highly significant.

As a crosscheck for the interpretation of this finding, we also investigated the distribution of TEWs against time,
which shows no such linear relation ($R=0.110$).
Therefore, we argue that the effect is not instrumental or caused by our data reduction, but physical.

  \subsection{Comparing high and low continuum level transits}
   \label{sec:HLContLevelTrans}
\begin{figure}[b]
     \centering
     \includegraphics[height=0.48\textwidth, angle=-90]{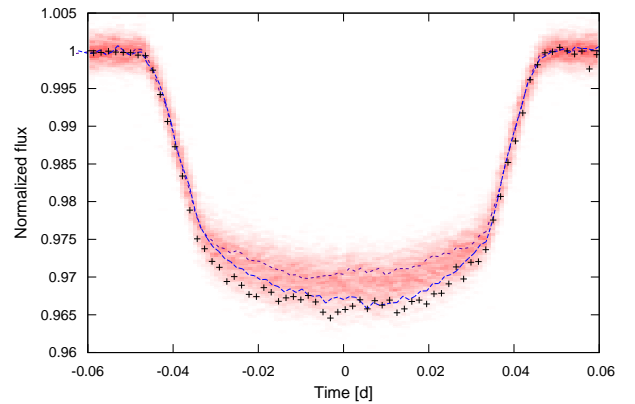}
     \caption{Average transit lightcurves obtained by combining the ten profiles which exhibit the highest (thick dashes) and
     lowest (thin dashes) continuum levels.
     The crosses indicate
     our lower envelope estimate and the color gradient (red) illustrates the distribution of data points for all
     available transits.
     \label{fig:TranProfs}}
   \end{figure}

   Since activity is evident in the profiles of the transit lightcurves, we further investigate its effect by
   comparing the most and least affected transit lightcurves. Therefore,
   we average the ten transits with the
   highest continuum levels (no. 3, 16, 42, 47, 50, 55, 68, 73, 76, and 81) and compare the result to an average of the ten transits
   with the lowest continuum level (no. 15, 23, 35, 40, 43, 69, 72, 75, 77, and 80). In Fig.~\ref{fig:TranProfs} we show
   the two averages as well as our computed lower envelope (see Sect.~\ref{sec:ObtLowEnv}) superimposed on the
   entire set of folded photometry data points. The distribution of the entire set is denoted by a color gradient (red) with
   stronger color indicating a stronger concentration of data points.
   The curve obtained from the
   transits at `low continuum state' is clearly shallower, as was already indicated by the TEW distribution presented in
   Fig.~\ref{fig:TEWvsCL}.
   
   The difference in TEW amounts to $\approx 15.5$~s in this extreme case.
   We checked the significance of this number with a Monte Carlo
   approach. On the basis of $20$ randomly chosen transits, we constructed two averaged lightcurves using $10$ transits for each
   and calculated the difference in TEW. Among $1\,000$ trials we did not find a single pair with a difference beyond $12$~s, so that
   the result is likely not due to an accidental coincidence.
   
  \subsection{Obtaining a lower envelope to the transit profiles}
   \label{sec:ObtLowEnv}
   As was demonstrated in the preceding section, activity shapes the transit lightcurves,
   and we cannot exclude that every transit is affected so that a priori no
   individual profile can be used as a template representing the `undisturbed' lightcurve. The
   distortion of the individual profiles is, however, not completely random, but the sign of the induced deviation is known as long as we assume
   that dark structures dominate over bright faculae which seems justified for CoRoT-2a \citep{Lanza2009}. In this case activity always tends to raise the 
   lightcurve level and, thus, decreases the transit depth.
   Therefore, the best model for the undisturbed profile can be obtained as a lower envelope to the observed
   transit profiles.
   
   Assume a set of $N_T$ transit observations, with the associated photometry folded at a single transit
   interval providing the set $LC_{T,i}$ of transit data points. If the lower envelope
   were observed (which might even be true), it would in principle look like every other lightcurve. In particular, it shows
   the same amount of intrinsic scattering (not including activity), characterized by the variance $\sigma_0^2$.
   
   Suppose we have an estimate of the variance
   \begin{equation}
     \sigma_0^2 \approx \frac{1}{N} \sum_j^N (LC_{T,j} - \mu_j)^2
     \label{eq:Variance}
   \end{equation}
   with the (unknown) expectation value $\mu_j$ and the number of data points $N$. The aim of the following effort is to identify the lowest
   conceivable curve sharing the same variance. In order to achieve this,
   we divide the transit span into a number of subintervals each containing a subsample, $s$, of $LC_{T}$.
   The distribution of data points in $s$ is now approximated by
   a `local model', $lm(\gamma)$, with a free normalization $\gamma$; $lm$ can for instance be a constant or
   a gradient. Given $lm$ we
   adapt the normalization in order to solve the equation
   \begin{equation}
     \left|\left(\frac{\sum_s (LC_s - lm(\gamma))^2 \cdot H(lm(\gamma)-LC_s)}{\sum_s H(lm(\gamma)-LC_s)} -\sigma_0^2 \right)\right| = 0
     \label{eq:VariEsti}
   \end{equation}
   where $H$ denotes the Heaviside function ($H(x)=1$ for $x>0$ and $H(x)=0$ otherwise). In this way we search for the local model compatible with the known variance of the lower
   envelope. The ratio on the
   left hand side of Eq.~\ref{eq:VariEsti} represents a variance estimator exclusively based on data points below the local model.
   It increases (strictly) monotonically except for the values of $\gamma$ where the local model
   `crosses' a data point and the denominator increases by one instantaneously. Therefore, there may be more than
   one solution to Eq.~\ref{eq:VariEsti}. From the mathematical point of view all solutions are equivalent, but for a conservative estimate
   of the lower envelope the largest one should be used.
   
   In Fig.~\ref{fig:TranProfs} we show the lower envelope which is in much better agreement with the average of
   the high continuum transit profiles than with its low continuum counterpart. The derivation of the lower envelope is based on
   Eq.~\ref{eq:VariEsti}. To obtain an estimate of $\sigma_0^2$ we fitted a $500$~s long span within the transit
   flanks ($3\,500\pm 250$~s from the transit center) where activity has little effect
   with a straight line and calculated the variance with respect to this model. The resulting value (using normalized flux)
   of $\sigma_0^2 = 1.6\times 10^{-6}$ was adopted in the calculation.
   Furthermore, we chose a bin width of $150$~s, and the `local model' was defined as a regression
   line within a $\pm 100$~s span around the bin center. Additionally, we postulated that
   at least $8$ (out of $\approx 350$) data points per bin should be located below the envelope,
   which proved to make the method more stable against the effect of outliers and has little impact otherwise.
   
  \subsection{Transit profiles in different color channels}
   CoRoT observes in three different bands termed `red', `green', and `blue'. In the following we present a qualitative analysis
   of the transit profiles in the separate bands.
   In the case of CoRoT-2a approximately $70$\%
   of the flux is observed in the red band, and the remaining $30$\% is more or less equally distributed among the green and
   blue channels. In order to compare the profiles we average all available
   transits in each band individually and normalize the results with respect to their TEW, i.e., after this step they all
   have the same TEW. The resulting profiles represent the curves which would be obtained if the stellar
   flux integrated along the planetary path was the same in all bands.
   
\begin{figure}[b]
     \centering
     \includegraphics[height=0.48\textwidth, angle=-90]{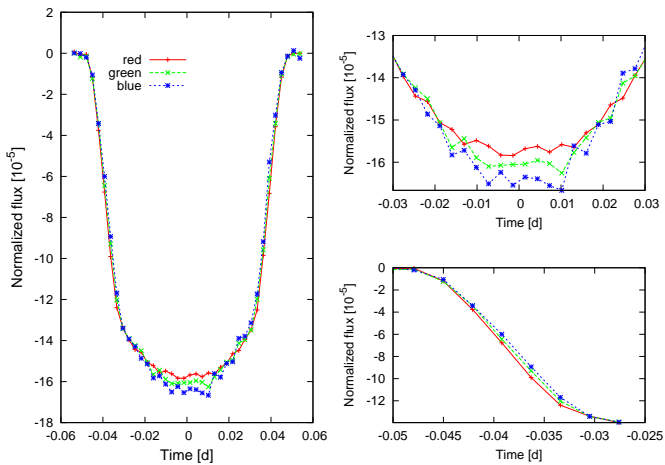}
     \caption{Left panel: Normalized transit in the three CoRoT bands red, green, and blue obtained by averaging all available data.
     Upper right: Close-up of the transit center. Lower right: Close-up of the ingress flank of the transit.
     \label{fig:ColorTran}}
   \end{figure}

   In Fig.~\ref{fig:ColorTran} we show the thus normalized
   transit lightcurves (TEW=1) obtained in the three bands.
   The normalized transits show a difference both in their flank profile and their depth. The blue and green transit profiles are
   narrower than the red one, and they are deeper in the center. This behavior is most pronounced in the blue band so that
   the green transit lightcurve virtually always lies in between the curves obtained in red and blue.
   
   The behavior described above can be explained by a color-dependent limb darkening law, with stronger limb
   darkening at shorter wavelengths as predicted by atmospheric models \citep{Claret2004} and observed on the Sun
   \citep[][]{Pierce1977}. Note that we checked
   that analytical transit models \citep{Pal2008} generated for a set
   of limb darkening coefficients, indeed, reproduce the observed behavior when normalized with respect to their TEW.
   
   Normalizing the averaged transits not with respect to TEW but using Eq.~\ref{eq:ziNorm} yields
   approximately the same depth in all bands, while the
   difference in the flanks becomes more pronounced. The reason for this could be a wrong relative normalization, which can
   e.g. occur if the eclipsed section of the star is (on average) redder than the rest of the surface due to
   pronounced activity or gravity darkening, or it may be a relic of an inappropriate treatment the companion's flux contribution.
   Whatever the explanation, it is clear from Fig.~\ref{fig:ColorTran} that the flanks
   and the centers in the individual bands cannot be reconciled simultaneously by a re-normalization.
   %because the offset changes its sign.
   Therefore, our analysis shows that the transit lightcurves are color dependent.
 
 \section{Stellar activity and planetary parameters}
   The preceding discussion shows that stellar activity does have a considerable influence on the
   profile of the transit lightcurves, and, therefore, the derivation
   of the planetary parameters will also be affected.
   Below we determine the radius and the orbit inclination of CoRoT-2b taking activity into account and discuss remaining
   uncertainties in the modeling.
 
 \subsection{Deriving the planetary radius and inclination from the lower envelope profile}
   \label{sec:DerivingPlaPar}
   In the analysis presented by \citet{Alonso2008}, the fit of the
   planetary parameters is based on the average of $78$~transit lightcurves (see Table~\ref{tab:Exo2prop} for an excerpt of their
   results). While this yields a good approximation,
   the results still include a contribution of stellar activity, and
   an undisturbed transit is needed to calculate `clean' planetary parameters.
   
   We follow a simplified approach to estimate the impact of activity on the planetary parameters. In particular, we use
   the lower envelope derived in Sect.~\ref{sec:HLContLevelTrans} as the best available model for the undisturbed transit.
   Starting from the
   results reported by \citet{Alonso2008}, we re-iterate the fit of the planetary parameters. In our approach we fix the parameters of
   transit timing, the semi-major axis and stellar radius, and the limb darkening coefficients at the values given by \citet{Alonso2008}
   (cf. Table~\ref{tab:Exo2prop}).
   The two free parameters are the planetary radius and its inclination.
   
   Note that limb darkening coefficients recovered
   by lightcurve analyses are not reliable especially when more than one coefficient is fitted \citep[e.g.,][]{Winn2009}.
   However, as an accurate calibration
   of the CoRoT color bands is not yet available and the coefficients determined by \citet{Alonso2008} roughly correspond to
   numbers predicted by stellar atmosphere models
   \footnote{For $T_{eff} = 5600$~K and $\log(g)=4.5$ the PHOENIX models given by \citet{Claret2004}
   yield quadratic limb darkening coefficients of $u_a=0.46$ and $u_b=0.25$ in the Sloan-$r'$ band.}, we
   decided to use the \citeauthor{Alonso2008} values here, too. This, furthermore, simplifies the comparison of the results.
   
   For the fit, we use the analytical models given by \citet{Pal2008} in combination with a Nelder-Mead simplex algorithm \citep[e.g.,][]{NR-BOOK}.

   \begin{figure}[h]
     \centering
     \includegraphics[height=0.48\textwidth, angle=-90]{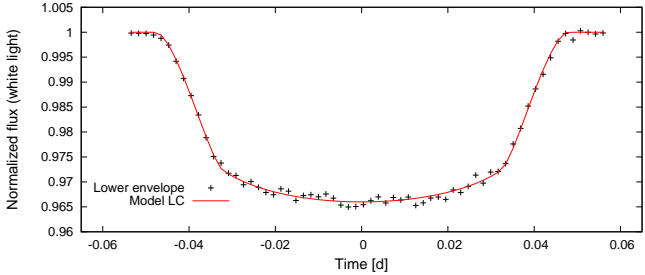}
     \caption{Lower envelope of all normalized transit lightcurves (already shown in Fig.~\ref{fig:TranProfs}) and our model fit.
     \label{fig:LowEnTran}}
   \end{figure}
   
   The result of our modeling is illustrated in Fig.~\ref{fig:LowEnTran}. The best radius ratio is
   $R_p/R_s = 0.172\pm 0.001$ at an inclination of \mbox{$87.7$\textdegree$\pm 0.2$\textdegree}. Note that the quoted errors are statistical errors,
   only valid in the context of the model. These numbers need to be compared to the values $R_p/R_s = 0.1667\pm 0.0006$  and
   \mbox{$87.84$\textdegree$\pm 0.1$\textdegree} (cf., Table~\ref{tab:Exo2prop}) derived without taking activity effects into account.
   The best fit inclination is compatible with the
   value determined by \citet{Alonso2008}, but `our' planet is larger by $\approx 3$\%.
   The planet's size mainly depends on the transit depth which is, indeed, affected at about
   this level by both normalization (Sect.~\ref{sec:QuanTheEffect}) and stellar activity (Sect.~\ref{sec:HLContLevelTrans}).
   
   Clearly, the derived change in $R_p/R_s$ of $0.005$ is much larger than the statistical error obtained from light curve
   fitting, and, therefore, the neglect of activity leads to systematic errors in excess of statistical errors. While the overall
   effect in planet radius is $\approx 3$\%, the error in density becomes $\approx 10$\%. Such errors are certainly
   tolerable for modeling planetary mass-radius relationships, they are not tolerable for precision measurements of possible
   orbit changes of such systems.
   
  \subsubsection{Planetary parameters and photospheric level}
   As already indicated the normalization according to Eq.~\ref{eq:ziNorm} relies on a `photospheric lightcurve level', $p$,
   which enters it as a global scaling factor and, therefore, also interferes with pinning down the planet's properties.
   
   In a simple case the star appears as a sphere with a purely photospheric surface, and the observed transit
   depth, $f_0$, can be identified with the squared ratio of planetary and stellar radii
   \begin{equation}
     f=\frac{\max(n_i-f_i)}{p} = \left( \frac{R_p}{R_*} \right)^2 L_d \, .
   \end{equation}
   Here $L_d$ is a correction factor accounting for limb darkening. Yet, when the observed star is active and the lightcurve
   is variable, there is no guarantee that the maximum point in the observed photometry is an appropriate representation
   of the photospheric stellar luminosity. Persistent inhomogeneities such as polar spots and long-lived spot contributions
   modulate the lightcurve, so that the pure photosphere might only be visible anytime the star is not observed or possibly
   never.
   
   Assume our estimate, $p_m$, of the photospheric level underestimates the `real' value, $p$, by a factor of
   $0 < c \le 1$ so that $p_m=p\cdot c$ and $f_{le,i}$ denotes the lower envelope transit lightcurve.
   Then the measured transit depth, $f_m$, becomes
   \begin{equation}
     f_m=\frac{\max(n_i-f_{le,i})}{p_m} = \left( \frac{R_p}{R_*} \right)^2 \frac{ L_d}{c} \, ,
     \label{eq:RpRs}
   \end{equation}
   and another scaling factor must be applied to the radius ratio. While $p_m$ is a measured quantity, $c$ is unknown, and
   if we neglect it in the physical interpretation, i.e., the right hand side of Eq.~\ref{eq:RpRs}, the ratio of planetary and
   stellar radii will be overestimated by a factor of $1/\sqrt{c}$.
   
   The value of $c$ cannot be quantified in the frame of this work; only an estimate can be provided. Doppler imaging studies
   revealed that polar spots are common, persistent structures in young, active stars \citep[e.g.][]{Huber2009}. Assuming
   that polar spots also exist on Corot-2a and that they reach down to a latitude of $70$\textdegree, they occupy roughly
   $2$\% of the visible stellar disk. Adopting a spot contrast of $50$\%,
   $c$ becomes $0.99$ in this case, and the planet size
   would be overestimated by $0.5$\%.
   As the poles of CoRoT-2a are seen under a large viewing  angle, their impact would, thus,
   be appreciably smaller than the amplitude of
   the global brightness modulation (ca. $4$\%).
   Nonetheless, in sign, this effect counteracts the transit depth decrease caused by activity, and
   if the polar spots are larger or symmetric structures at lower latitudes contribute, it may even balance it.
   
 \section{Discussion and conclusion}
   Stellar activity is clearly seen in the CoRoT measured transit lightcurves of CoRoT-2a, and an
   appropriate normalization is necessary to preserve as much of the true transit lightcurve profile as possible.
   
   The transit profiles observed in CoRoT-2a are affected by activity, as is obvious in many transits where active regions
   cause distinct `bumps' in the lightcurve \citep[e.g.][]{Wolter2009}. Yet, our
   analysis reveals that not only profiles with bumps but presumably all transit profiles are influenced by
   stellar activity. This is evident in a
   relationship between the transit equivalent width and the level of the global lightcurve: Transits observed during
   periods where the star appears relatively bright are deeper than those observed during faint phases.
   We demonstrated this correlation to be extremely significant, but also the data points show a large scatter around an
   assumed linear model relation.
   If the star were to modulate its surface brightness globally and homogeneously, this relation would be perfectly
   linear except for measurement errors. Therefore, we interpret the observed scatter as a consequence of surface
   evolution. When the global lightcurve is minimal, we also find more spots under the eclipsed portion of the surface, but only
   on average and, for an individual transit, this needs not to be the case.  Thus, the surface configuration is clearly not the same
   for every minimum observed.
   
   Additionally, we demonstrated that the transit profiles show a color dependence compatible with a color-dependent limb darkening
   law as expected from stellar atmospheric models and the solar analogy.
   
   All these influences potentially interfere with the determination of the planetary parameters.
   Using our lower (white light) transit envelope, we determined new values for the planet-to-star radius ratio and the orbital inclination.
   While the latter remains compatible with previously reported results, the planet radius turns out to be larger (compared to the star)
   by about $3$\%.
   Although our approach accounts for many activity related effects, a number of uncertainties remain.
   For example the photospheric lightcurve level needed for transit normalization cannot be determined with
   certainty from our analysis and the same applies to the limb darkening law.
   More certain than the planetary parameters themselves is, therefore, our conclusion that the errors in their determination
   are much larger than the statistical ones.
   
   While CoRoT-2a is certainly an extreme example of an active star, stellar activity is a common phenomenon especially on
   young stars. Therefore, in general,
   stellar activity cannot be neglected in planetary research if the accuracy of
   the results is to exceed the percent level.

%%%%%%%%%%%%%%%%%%%%%%%%%%%%%%%%%%%%%%%%%%%%%%%%%%%%%%%%%%%
\begin{acknowledgements}
S.C. and U.W. acknowledge DLR support (50OR0105).
K.H. is a member of the DFG Graduiertenkolleg 1351 \textit{Extrasolar Planets and their Host Stars}.
S.S. acknowledges DLR support (50OR0703).
\end{acknowledgements}

%%%%%%%%%%%%%%%%%%%%%%%%%%%%%%%%%%%%%%%%%%%%%%%%%%%%%%%%%%%%%%

\bibliographystyle{aa}
\bibliography{referenz}

\begin{thebibliography}{12}
\expandafter\ifx\csname natexlab\endcsname\relax\def\natexlab#1{#1}\fi

\bibitem[{{Alonso} {et~al.}(2008){Alonso}, {Auvergne}, {Baglin}, {Ollivier},
  {Moutou}, {Rouan}, {Deeg}, {Aigrain}, {Almenara}, {Barbieri}, {Barge},
  {Benz}, {Bord{\'e}}, {Bouchy}, {de La Reza}, {Deleuil}, {Dvorak}, {Erikson},
  {Fridlund}, {Gillon}, {Gondoin}, {Guillot}, {Hatzes}, {H{\'e}brard},
  {Kabath}, {Jorda}, {Lammer}, {L{\'e}ger}, {Llebaria}, {Loeillet}, {Magain},
  {Mayor}, {Mazeh}, {P{\"a}tzold}, {Pepe}, {Pont}, {Queloz}, {Rauer},
  {Shporer}, {Schneider}, {Stecklum}, {Udry}, \& {Wuchterl}}]{Alonso2008}
{Alonso}, R., {Auvergne}, M., {Baglin}, A., {et~al.} 2008, \aap, 482, L21

\bibitem[{{Auvergne} {et~al.}(2009){Auvergne}, {Bodin}, {Boisnard}, {Buey},
  {Chaintreuil}, \& {CoRoT team}}]{Auvergne2009}
{Auvergne}, M., {Bodin}, P., {Boisnard}, L., {et~al.} 2009, ArXiv e-prints

\bibitem[{{Bouchy} {et~al.}(2008){Bouchy}, {Queloz}, {Deleuil}, {Loeillet},
  {Hatzes}, {Aigrain}, {Alonso}, {Auvergne}, {Baglin}, {Barge}, {Benz},
  {Bord{\'e}}, {Deeg}, {de La Reza}, {Dvorak}, {Erikson}, {Fridlund},
  {Gondoin}, {Guillot}, {H{\'e}brard}, {Jorda}, {Lammer}, {L{\'e}ger},
  {Llebaria}, {Magain}, {Mayor}, {Moutou}, {Ollivier}, {P{\"a}tzold}, {Pepe},
  {Pont}, {Rauer}, {Rouan}, {Schneider}, {Triaud}, {Udry}, \&
  {Wuchterl}}]{Bouchy2008}
{Bouchy}, F., {Queloz}, D., {Deleuil}, M., {et~al.} 2008, \aap, 482, L25

\bibitem[{{Claret}(2004)}]{Claret2004}
{Claret}, A. 2004, \aap, 428, 1001

\bibitem[{{Huber} {et~al.}(2009){Huber}, {Wolter}, {Czesla}, {Schmitt},
  {Esposito}, {Ilyin}, \& {Gonz{\'a}lez-P{\'e}rez}}]{Huber2009}
{Huber}, K.~F., {Wolter}, U., {Czesla}, S., {et~al.} 2009, ArXiv e-prints,
  accepted by \aap

\bibitem[{{Lanza} {et~al.}(2009){Lanza}, {Pagano}, {Leto}, {Messina},
  {Aigrain}, {Alonso}, {Auvergne}, {Baglin}, {Barge}, {Bonomo}, {Boumier},
  {Collier Cameron}, {Comparato}, {Cutispoto}, {de Medeiros}, {Foing},
  {Kaiser}, {Moutou}, {Parihar}, {Silva-Valio}, \& {Weiss}}]{Lanza2009}
{Lanza}, A.~F., {Pagano}, I., {Leto}, G., {et~al.} 2009, \aap, 493, 193

\bibitem[{{P{\'a}l}(2008)}]{Pal2008}
{P{\'a}l}, A. 2008, \mnras, 390, 281

\bibitem[{{Pierce} \& {Slaughter}(1977)}]{Pierce1977}
{Pierce}, A.~K. \& {Slaughter}, C.~D. 1977, \solphys, 51, 25

\bibitem[{{Pont} {et~al.}(2007){Pont}, {Gilliland}, {Moutou}, {Charbonneau},
  {Bouchy}, {Brown}, {Mayor}, {Queloz}, {Santos}, \& {Udry}}]{Pont2007}
{Pont}, F., {Gilliland}, R.~L., {Moutou}, C., {et~al.} 2007, \aap, 476, 1347

\bibitem[{{Press}(1992)}]{NR-BOOK}
{Press}, W.~H. 1992, {Numerical Recipes in C, The Art of Scientific Computing},
  2nd edn. (New York: Cambridge University Press)

\bibitem[{{Winn}(2009)}]{Winn2009}
{Winn}, J.~N. 2009, in IAU Symposium, Vol. 253, IAU Symposium, 99--109

\bibitem[{{Wolter} {et~al.}(2009){Wolter}, {Schmitt}, {Huber}, {Czesla},
  {M\"uller}, {Guenther}, \& {Hatzes}}]{Wolter2009}
{Wolter}, U., {Schmitt}, J.~H.~M.~M., {Huber}, K.~F., {et~al.} 2009, {accepted
  by \aap}

\end{thebibliography}

%%%%%%%%%%%%%%%%%%%%%%%%%%%%%%%%%%%%%%%%%%%%%%%%%%%%%%%%%%%%%%%
%\appendix

\end{document}